\documentclass[conference]{IEEEtran}

  	\usepackage[pdftex]{graphicx}
  	\graphicspath{{../pdf/}{../jpeg/}}
	\DeclareGraphicsExtensions{.pdf,.jpeg,.png}

	\usepackage[cmex10]{amsmath}
	\usepackage{mathabx}
	\usepackage{algorithmic}
	\usepackage{array}
	\usepackage{mdwmath}
	\usepackage{mdwtab}
	\usepackage{eqparbox}
	\usepackage{url}
\usepackage{xcolor}
\usepackage{dblfloatfix}
\usepackage{wrapfig}

\begin{document}
\title{Large Language Model-Driven Cross-Domain Orchestration Using Multi-Agent Workflow}

\author{\authorblockN{
    Xiaonan~Xu\authorrefmark{1}, 
    Haoshuo~Chen\authorrefmark{1}, 
    Jesse~E.~Simsarian\authorrefmark{1},  
    Roland~Ryf\authorrefmark{1},  
    Nicolas~K.~Fontaine\authorrefmark{1},  \\   
    Mikael~Mazur\authorrefmark{1}, 
    Lauren~Dallachiesa\authorrefmark{1},  
    David~T.~Neilson\authorrefmark{1}}

 \authorblockA{\authorrefmark{1}Nokia Bell Labs, 600 Mountain Ave., Murray Hill, NJ 07974, USA \\xuxiaonan2008@gmail.com}
}
\maketitle
\vspace{-20pt}
\begin{abstract}
    We showcase an application that leverages multiple agents, powered by large language models and integrated tools, to collaboratively solve complex network operation tasks across various domains. 
    The tasks include real-time topology retrieval, network optimization using physical models, and fiber switching facilitated by a robotic arm.
\end{abstract}

\IEEEoverridecommandlockouts

\IEEEpeerreviewmaketitle

\section{Introduction}

Attention is increasingly focused on integrating AI, including large language models (LLMs), into domain-specific applications\cite{10484825,huang2023large}. 
However, the use of LLMs for orchestrating networks across multiple domains has not yet been demonstrated.
Cross-domain network orchestration is essential for delivering dynamic, scalable, and high-performance services in today's vertical networks\cite{cross_domain}.

In this paper, we demonstrate an orchestration application over a robotic domain and an optical transport network (OTN) domain using multi-agent conversation~\cite{wu2023autogen}. 
This orchestration application consists of multiple interacting LLM-empowered agents that exchange information and collaborate to solve problems beyond the capability of individual agents.
Orchestration across multiple domains is achieved through two key features of intelligent agents.
First, they can access and utilize a variety of domain-specific management interfaces, tools, and functionalities through function calling.
This enables, for example, retrieving real-time network parameters, using models to analyze network Quality of Transmission (QoT), and planning collision-free robot motion~\cite{xu2023automated}.
Second, LLM empowers agents not only to understand and generate human-like text, including code blocks, but also to facilitate efficient communication between agents to complete complex tasks.
This capability enables, for instance, the suggestion of detailed plans for complex tasks and the selection of the most suitable agent for each specific subtask.

\begin{figure}[b]
\vspace{-14pt}
  \begin{minipage}[t]{1\linewidth}
    \includegraphics[width=\linewidth]{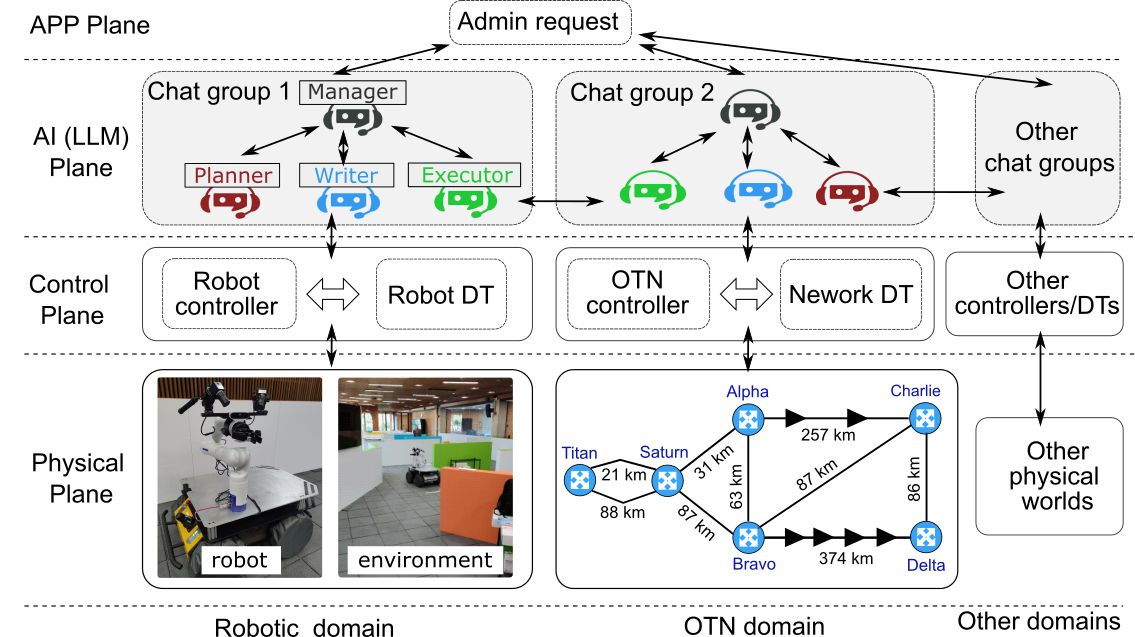}
    \vspace{-22pt}
    \caption{Architecture and workflow of cross-domain network automation enabled by LLM-empowered agents. (DT: digital twin)}
    \label{fig:arch}
  \end{minipage}\rule{1em}{0pt}%
\end{figure}

\section{Application Description}

Figure~\ref{fig:arch} illustrates the orchestration architecture and workflow of network automation, spanning the robotic and OTN domains.
In the network domain, we utilize a laboratory optical network comprising six commercial Nokia 1830 Photonic Service Switch (PSS) transport nodes equipped with flexgrid reconfigurable optical add-drop multiplexers (ROADMs), with fiber lengths indicated in the network topology. 
An OTN controller can collect and analyze real-time network data and employ network digital twins with the QoT estimator GNPy~\cite{ferrari2020gnpy} for route planning and optimization.
In the robotic domain, a real robotic arm with a 2-finger gripper mounted on a mobile robotic base with Lidar is applied to perform physical tasks in the laboratory~\cite{xu2023automated}.
The robot controller manages and controls the robot's operations and can interact with the robot's digital twin, like MoveIt~\cite{moveit}, for movement planning.
The orchestration architecture can be extended to support more network domains such as 5G network, and other emerging technologies such as augmented/virtual reality (AR/VR)~\cite{chen2023digital}.

We assign one chat group to the OTN domain and another to the robotic domain, each containing multiple intelligent agents with specific roles:
1) a manager automatically selects the appropriate agent based on the task request, each agent's responsibilities, and their responses, 
2) a planner suggests a plan to accomplish the request, 
3) a writer handles language tasks such as retrieving data and generating code, with access to device controllers and various tools and 
4) an executor runs code contained in input messages and sends out result messages.
The agents within each chat group can communicate freely, and different groups are linked through the executors.

\begin{figure*}[t]
   \centering
        \includegraphics[width=18cm]{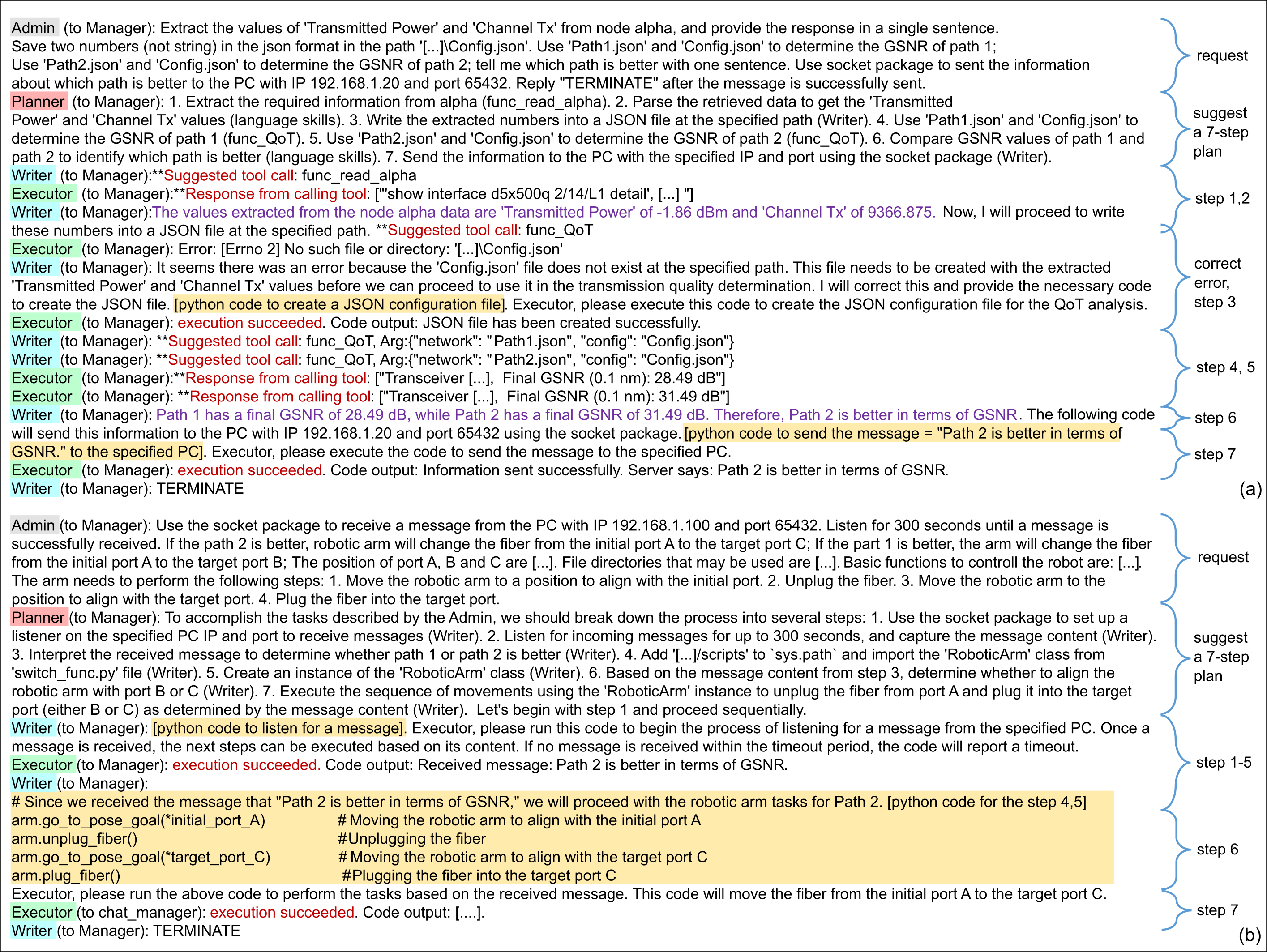}
        \vspace{-2pt}
    \caption{Output of multi-agent conversation involving a\colorbox{red!30}{Planner},\colorbox{cyan!30}{Writer}, and\colorbox{green!30}{Executor} from the OTN domain (a) and the robotic domain (b), respectively.
    The\colorbox{black!20}{Admin}inputs the task request. 
    Responses highlighted in \textcolor{red}{red} signify the usage of tools such as function calling and code execution. 
    Responses highlighted in \textcolor{violet}{purple} exemplify the usage of the LLM's language skills such as data retrieval and simple mathematical reasoning.
    "[...]" indicates omitted information such as paths for saving files, code output, and descriptions of pre-defined functions.
    The\colorbox{yellow!30}{yellow box}signifies omitted code or part of code generation.
    }
    \label{fig:output}
\vspace{-16pt}
\end{figure*}

We demonstrate the interactive multi-agent conversations in which 1) the network domain evaluates the generalized signal-to-noise ratio (GSNR) of two paths based on retrieved data from the real network, determines a better path, and shares the result with the robotic domain, and 
2) the robotic domain listens and receives the message, sends specific commands to the robot, and executes the physical action of switching the fiber.
Figure~\ref{fig:output}(a) and (b) exemplify the dialogues of the OTN and the robotic chat group, respectively.
Upon receiving the request from an administrator, the planner divides the task and suggests a plan with multiple steps.
The writer and the executor utilize language skills, pre-defined functions, or code generation to accomplish sub-tasks step-by-step.
The agents can fix code and analyze the problem if an error is indicated, as shown before step 3 in Figure~\ref{fig:output}(a).
Step 6 in Figure~\ref{fig:output}(b) shows the writer generating code using basic pre-defined functions to operate the robot, instructing it to unplug the fiber from port A and plug it into port C after receiving the message from the OTN domain.
All the agents are configured using GPT-4.

Future work will focus on extending the application to support more network domains and utilizing open-sourced and local fine-tuned LLMs for enhanced data security, domain-specific tuning, and reduced latency.

\vspace{-2pt}
\bibliographystyle{IEEEtran}
\bibliography{IEEEabrv,biblio_traps_dynamics}

\end{document}